# Quantum Information of a Four-Level Tripod-Type Atom in Motion Interacting with a Deformed Binomial Field in the Presence of a Non-Linear Medium

*Sameh. T. Korashy\**

*\* Department of Basic Sciences, Technology Financial & Management Higher Institute, Sohag, Egypt*

**Abstract** This paper investigates the interaction dynamics of a four-level tripod-type atomic system coupled to a q-deformed binomial field state within a Kerr-medium. The interaction model incorporates time-dependent coupling parameter and detuning parameter, providing a more adaptable framework for describing atom-field interactions. Special focus is placed on examining how the q-deformation, time-dependent coupling parameter, detuning parameter and Kerr nonlinearity affect the system's fidelity properties and linear entropy dynamics. Our results demonstrate that the effects of the considered parameters have a significant impact on atom-field entanglement and fidelity. These findings offer valuable insights into controlled quantum systems, with potential applications in quantum information processing and nonlinear quantum optics.

# 1 Introduction

The study of atom-field interactions forms a cornerstone in the domain of quantum optics and quantum information [1-3]. The interaction between atoms and electromagnetic fields underpins many quantum phenomena and is critical in understanding the evolution of quantum systems. The theoretical foundation of atom-field interactions was laid in the early 20th century with the development of quantum mechanics. The seminal works of Dirac [4] provided a basis for understanding light-matter interaction at the quantum level. The Jaynes-Cummings model (JCM) [5], introduced in 1963, marked a major milestone by offering a simplified yet highly insightful representation of the coupling between a two-level atom and a single quantized mode of the electromagnetic field. Extensions of the JCM to multi-level atoms [6-11] and multi-mode fields [12-15] have since expanded its applicability.

One of the significant outcomes of studying atom-field interactions is the understanding and quantification of quantum entanglement. Measures such as von Neumann entropy [3, 16], linear entropy [3, 17-19], and fidelity [3, 20] have been extensively used to evaluate the degree of entanglement and coherence in quantum systems. Von Neumann entropy, for instance, provides a measure of the information content of a quantum state and has been applied to study decoherence in atom-field interactions [21-23]. Linear entropy offers a simpler computational alternative to von Neumann entropy and has been employed to investigate mixedness and purity in evolving quantum states [24, 25]. Moreover, the concept of mixedness, which quantifies the departure of a quantum state from being pure, has been critical in understanding the effects of environmental noise on quantum systems [26,27].

Modern studies have significantly advanced the understanding of time-dependent interactions, offering new methodologies to solve the Schrodinger equation under time-varying coupling parameters [28-32]. Such approaches are particularly relevant for controlling quantum systems in dynamic environments, such as in quantum gates and sensors. Recent advancements have also highlighted the importance of fidelity, a measure of the closeness between quantum states, in evaluating the robustness of quantum operations and protocols [33, 34].

On the other hand, the study of non-classical states such as binomial states [35, 36] and coherent states [37,38] has been pivotal in exploring quantum optics and quantum information, especially in the context of atom-field interactions. Recently, q-deformed versions of these states namely q-deformed binomial states [39, 40] and q-deformed coherent states [41] have attracted attention due to their ability to generalize traditional quantum mechanical states within the framework of q-deformation theory [42-44], allowing for a richer structure of quantum superposition and interaction properties.

In this study, we examine the time-dependent interaction between a four-level tripod-type atom and a single mode q-deformed binomial field, in which the q-deformed binomial state generalizes classical binomial distributions into quantum mechanical regime, displaying non-classical characteristics influenced by the deformation parameter q. The paper is structured as follows: In Sec. 2, we present the model and derive its solution under a specific approximation akin to the Rotating-Wave Approximation (RWA) for all times $t > 0$. Sec. 3 focuses on analyzing the fidelity and dynamical properties across various regimes. Numerical results related to linear entropy are discussed in Sec. 4. Finally, the main findings and conclusions are summarized in Sec. 5.

# 2 Physical model

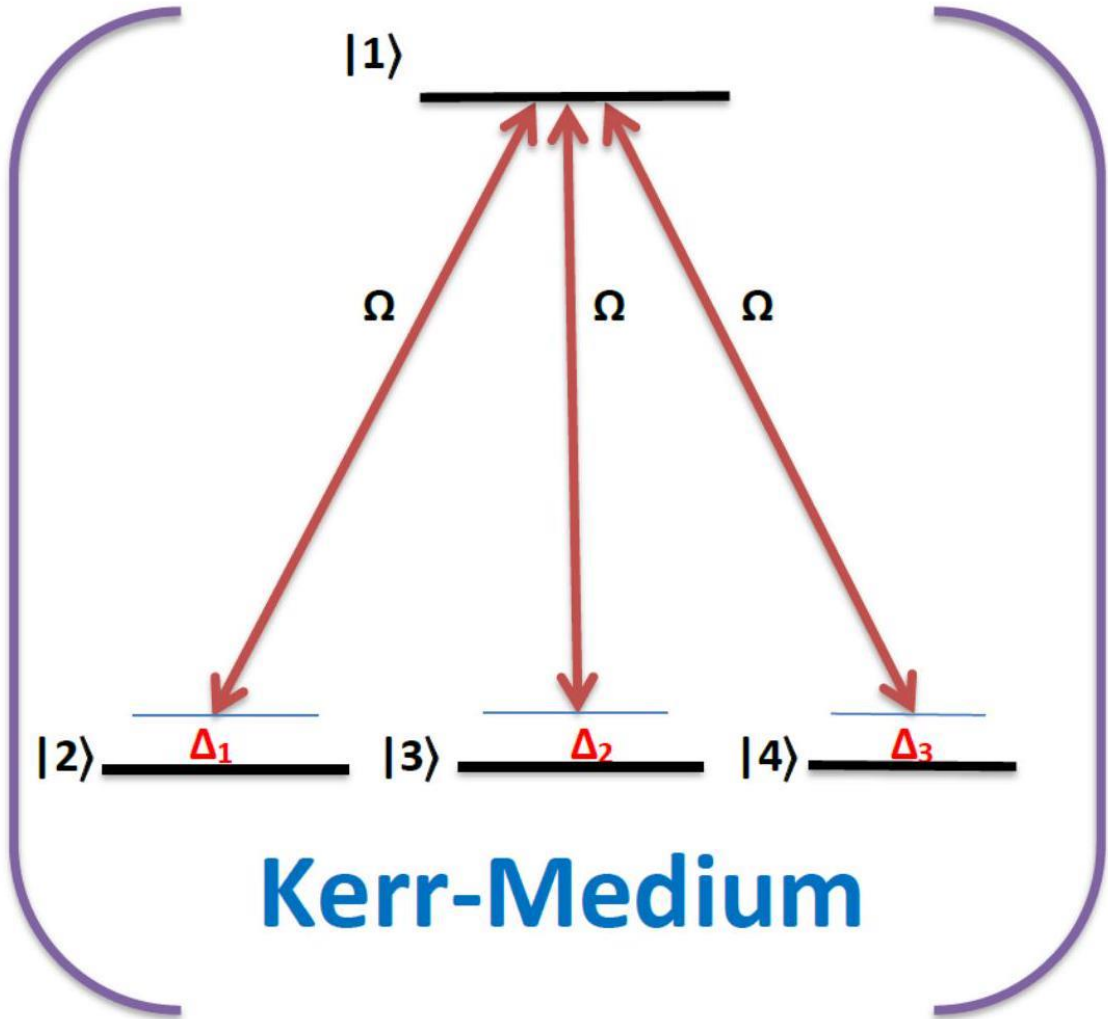

Fig. 1 Schematic diagram of a moving four-level tripod-type atom interacting with a single mode q-deformed binomial field.

The considered physical model is a time-dependent system consisting of a Four-level tripod-type atom interacting with a single mode deformed binomial field within an optical cavity, surrounded by Kerr nonlinearity and influenced by detuning parameter. The four atomic levels are respectively labeled as $|j\rangle, j = 1,2,3,4$, with corresponding energies $\omega_j$, and the field has a frequency $\Omega$. The transitions $|1\rangle \leftrightarrow |2\rangle, |1\rangle \leftrightarrow |3\rangle$, and $|1\rangle \leftrightarrow |4\rangle$ are allowed as depicted in Fig. 1. In the Rotating-Wave Approximation (RWA), the interaction Hamiltonian for the described physical system is given by ($\hbar = 1$):

$$\hat{H}_I = \sum_{r=1}^{3} \kappa_r(t)\left(\hat{a}_q e^{-i\Delta_r t}\hat{S}_{1r+1}\right) + h.c. + \chi \hat{a}_q^{\dagger 2} \hat{a}_q^2 \qquad (2.1)$$

In which, the operators $\hat{S}_{ij} = |i\rangle\langle j|$ represent the atomic raising or lowering operators, while $\hat{a}_q^{\dagger}(\hat{a}_q)$, are the field creation and annihilation operators of the deformed field mode, respectively. The parameters $\kappa_i(t), i = 1,2,3$ correspond to the atom-field coupling parameters, and $\chi$ denotes the third-order nonlinearity of the Kerr-medium. The detuning parameters $\Delta_1, \Delta_2, \Delta_3$ are defined as follows:

$$\Delta_r = \omega_{r+1} - \omega_1 + \Omega, r = 1,2,3 \tag{2.2}$$

We define $\kappa_1(t) = \kappa_2(t) = \kappa_3(t) = \kappa(t) = \lambda_j \cos(\mu t) = \frac{\lambda_j}{2}(e^{i\mu t} + e^{-i\mu t})$, where, $\lambda_j, \mu, j = 1,2,3$ are arbitrary constants. As observed, the Hamiltonian contains two exponential terms: one with rapid oscillations, $e^{\pm i(\Delta_j + \mu)t}$, and another with slower variations, $e^{\pm i(\Delta_j - \mu)t}$. In this case, if we neglect the rapidly oscillating terms in comparison to the slowly varying ones, the interaction Hamiltonian can be rewritten in the following manner

$$\hat{H}_I = \sum_{r=1}^{3} \frac{\lambda_r}{2} (\hat{a}_q e^{-i\epsilon_r t} \hat{S}_{1r+1}) + \text{h.c.} + \chi \hat{a}_q^{\dagger 2} \hat{a}_q^2$$

where,

$$\epsilon_r = \Delta_r - \mu, r = 1,2,3 \tag{2.3}$$

As an application of the q-deformed binomial distribution, we construct the q-deformed binomial state $|\tau, M\rangle_q$ for a single-mode of the q-boson as follows [39, 40]:

$$|\tau, M\rangle_q = \sum_{n=0}^{M} \beta_n^M |n\rangle_q, \beta_n^M = \sqrt{b(n; M, \tau)} \tag{2.4}$$

where $|n\rangle_q$ is the q-number state and the q-deformed binomial distribution $b(n; M, \tau)$ is defined as

$$b(n; M, \tau) = \binom{M}{n}_q \tau^n (1-\tau)_q^{M-n}, 0 < \tau < 1 \tag{2.5}$$

where,

$$\binom{M}{n}_q = \frac{[M]_q!}{[M-n]_q! [n]_q!},$$

$$[n]_q! = [n]_q [n-1]_q \ldots [1]_q, [0]_q! = 1, [n]_q = \frac{1 - q^{-n}}{1 - q}$$

Using the q-binomial theorem [45-47]

$$(1-\tau)_q^{M-n} = \sum_{k=0}^{M-n} \binom{M-n}{k}_q q^{k(k-1)/2} (-\tau)^k, 0 < q < 1.$$

When applying the operators $\hat{a}_q$ and $\hat{a}_q^{\dagger}$ on $|\tau, M\rangle_q$, then we have

$$\hat{a}_q|\tau, M\rangle_q = \sqrt{b(n; M, \tau)[n]_q}|n-1\rangle_q$$

$$\hat{a}_q^\dagger|\tau, M\rangle_q = \sqrt{b(n; M, \tau)[n+1]_q}|n+1\rangle_q$$

Now, we assume that the wave function of the atom-field at any time $t > 0$ can be expressed as:

$$|\Psi(t)\rangle = \sum_{n=0}^{M} \beta_n^M \big[\psi_1(n,t)|n,1\rangle_q + \psi_2(n+1,t)|n+1,2\rangle_q$$

To achieve this, suppose that the atom-field initial state is

$$|\Psi(0)\rangle = \sum_{n=0}^{M} \beta_n^M \big[\vartheta_1|n,1\rangle_q + \vartheta_2|n+1,2\rangle_q + \vartheta_3|n+1,3\rangle_q + \vartheta_4|n+1,4\rangle_q\big] \qquad (2.7)$$

by substituting $|\Psi(t)\rangle$ from Eq. (2.7) and $\hat{H}_I$ from Eq. (2.3) into the time-dependent Schrodinger equation $i\frac{\partial}{\partial t}|\Psi(t)\rangle = \hat{H}_I|\Psi(t)\rangle$, we derive the following coupled differential equations for the atomic probability amplitudes

$$i\frac{d}{dt}\begin{pmatrix}\psi_1(n,t)\\ \psi_2(n+1,t)\\ \psi_3(n+1,t)\\ \psi_4(n+1,t)\end{pmatrix} = \begin{pmatrix} v_1 & g_1 e^{-i\epsilon_1 t} & g_2 e^{-i\epsilon_2 t} & g_3 e^{-i\epsilon_3 t}\\ g_1 e^{-i\epsilon_1 t} & v_2 & 0 & 0\\ g_2 e^{i\epsilon_2 t} & 0 & v_3 & 0\\ g_3 e^{-i\epsilon_3 t} & 0 & 0 & v_4 \end{pmatrix}\begin{pmatrix}\psi_1(n,t)\\ \psi_2(n+1,t)\\ \psi_3(n+1,t)\\ \psi_4(n+1,t)\end{pmatrix} \qquad (2.8)$$

where

$$v_1 = \chi[n]_q[n-1]_q, v_2 = v_3 = v_4 = \chi[n]_q[n+1]_q$$

$$g_1 = \frac{\lambda_1}{2}\sqrt{[n+1]_q}, g_2 = \frac{\lambda_2}{2}\sqrt{[n+1]_q},$$

Next, under the assumption that $\omega_2 = \omega_3 = \omega_4$ and thus $\Delta_1 = \Delta_2 = \Delta_3 = \Delta$, Also, we can also assume $\lambda_1 = \lambda_2 = \lambda_3 = \lambda$, which results in $g_1 = g_2 = g_3 = g$. Under these conditions, the probability amplitudes $\psi_2(n+1,t), \psi_3(n+1,t)$ and $\psi_4(n+1,t)$ satisfy similar differential equations, leading to the conclusion that $\psi_2(n+1,t) = \psi_3(n+1,t) = \psi_4(n+1,t)$. Thus, the atom-field system admits an analytical solution, reducing the four-level system to a two-level syatem. Consequently, the coupled system of differential equations for the probability amplitudes in Eq. (2.9) can be simplified to:

$$i\frac{d}{dt}\begin{pmatrix}\psi_1(n,1)\\ \psi_2(n+1,2)\end{pmatrix} = \begin{pmatrix} v_1 & 3ge^{-i\epsilon t}\\ ge^{-i\epsilon t} & v_2 \end{pmatrix}\begin{pmatrix}\psi_1(n,1)\\ \psi_2(n+1,2)\end{pmatrix}. \qquad (2.10)$$

The solution of Eqs. (2.9) are given as follows:

$$\psi_1(n,t) = \sum_{j=1}^{2} B_j e^{iX_j t}$$

By applying these initial conditions for atom and field, and using Eq. (2.12), the coefficients $B_j$ are determined as:

$$B_j = \frac{-(X_k + v_1)\vartheta_1 - 3g\vartheta_2}{X_j - X_k}, j \neq k = 1,2 \tag{2.12}$$

where $X_j, j = 1,2$ are the roots of the equation

$$X^2 + a_1 X + a_2 = 0 \tag{2.13}$$

with

$$a_1 = \epsilon + v_1 + v_2, a_2 = v_1(\epsilon + v_2) - 3g^2.$$

For any time $t > 0$ the reduced density matrix of the atom is given by:

$$\hat{\varrho}_A(t) = \text{Tr}_F(|\Psi(t)\rangle\langle\psi(t)|) = \begin{pmatrix} \varrho_{11}(t) & \varrho_{12}(t) & \varrho_{13}(t) & \varrho_{14}(t) \\ \varrho_{21}(t) & \varrho_{22}(t) & \varrho_{23}(t) & \varrho_{24}(t) \\ \varrho_{31}(t) & \varrho_{32}(t) & \varrho_{33}(t) & \varrho_{34}(t) \\ \varrho_{41}(t) & \varrho_{42}(t) & \varrho_{43}(t) & \varrho_{44}(t) \end{pmatrix} \tag{2.14}$$

where

$$\varrho_{11}(t) = \sum_{n=0}^{M} |\beta_n^M|^2 \psi_1(n,t)\psi_1^*(n,t)$$

$$\varrho_{22}(t) = \sum_{n=0}^{M} |\beta_n^M|^2 \psi_2(n+1,t)\psi_2^*(n+1,t), \ldots\ldots$$

$$\varrho_{22}(t) = \varrho_{33}(t) = \varrho_{44}(t) = \varrho_{23}(t) = \varrho_{24}(t) = \varrho_{34}(t),$$

The reduced density operator of the field $\hat{\varrho}_F(t)$ is given by

$$\hat{\varrho}_F(t) = \text{Tr}_A(|\Psi(t)\rangle\langle\psi(t)|). \tag{2.16}$$

In the subsequent sections, for simplicity, we assume that the constants $\lambda_i = \lambda$ are real and the interaction time is the scaled as $T = \lambda t$.

# 3 Fidelity

It is important to highlight that the collapse and revival phenomenon provides insight into the behavior of the atom-field interaction. Therefore, we will examine the dynamics of a key quantity, specifically fidelity. When the input state is a pure state, denoted as $\rho_{in} = |\Psi(0)\rangle\langle\Psi(0)|$, the fidelity is defined as the quantum overlap between the input and output states. In this case, the fidelity is given by [3, 20].

$$F(t) = \langle\Psi(t)|\rho_{in}|\Psi(t)\rangle = |\langle\Psi(t) | \Psi(0)\rangle|^2. \tag{3.1}$$

Using Eqs. (2.7), (2.8), and (3.1), we obtain

$$F(t) = \left| \sum_{n=0}^{M} [\beta_n^M \alpha_1 \psi_1^*(n,t) + 3\beta_{n+1}^M \alpha_2 \psi_2^*(n+1,t)] \right|^2$$

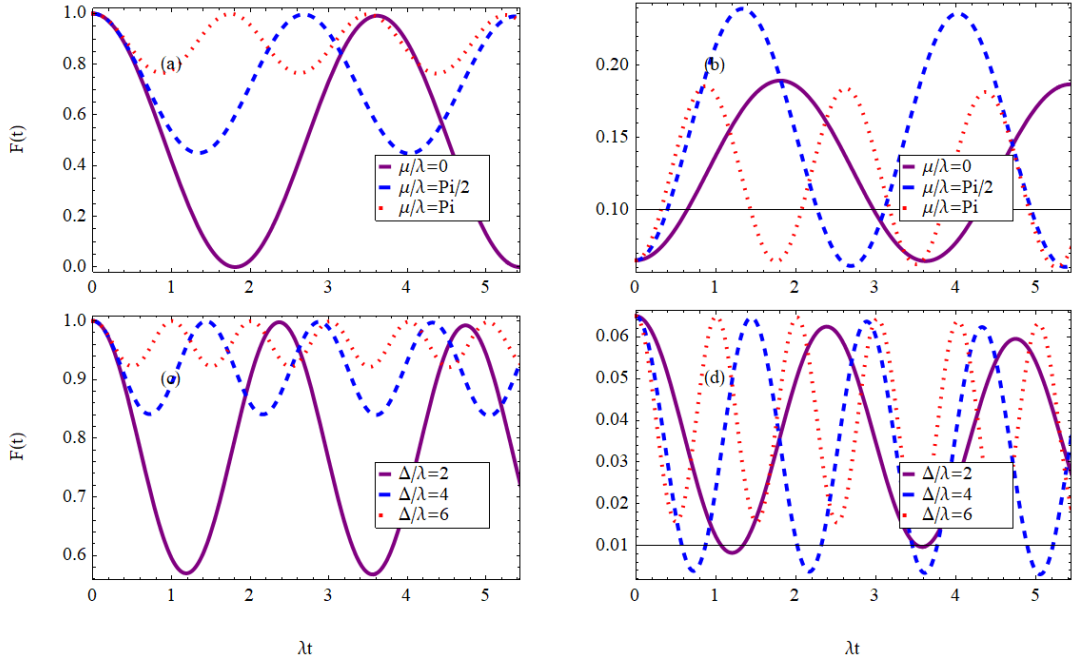

Fig. 2 Evolution of the fidelity $F(T)$ for a four-level atom interacting with a single mode deformed binomial field for $\alpha_1 = 1, \alpha_2 = \alpha_3 = \alpha_4 = 0$ (left plot), and for $\alpha_1 = \alpha_2 = \alpha_3 = \alpha_4 = 0.5$ (right plot) for the parameters $q = 0.9, \tau = 0.0007$, and for: (a, b) $\mu/\lambda = 0, \mu/\lambda = Pi/2, \mu/\lambda = Pi$, (c, d) $\Delta/\lambda = 2, \Delta/\lambda = 4, \Delta/\lambda = 6$.

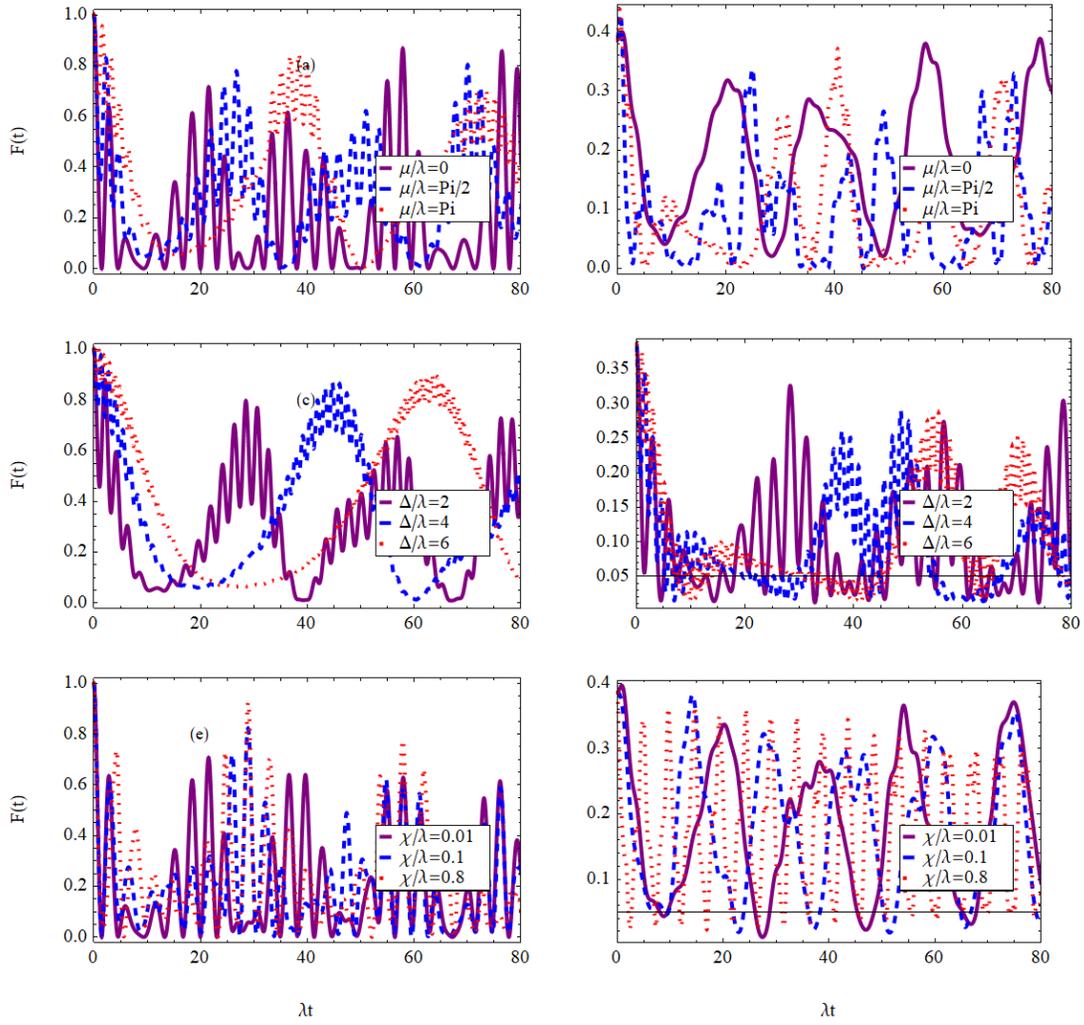

Fig. 3 Evolution of the fidelity $F(T)$ for a four-level atom interacting with a single mode deformed binomial field for $\alpha_1 = 1, \alpha_2 = \alpha_3 = \alpha_4 = 0$ (left plot), and for $\alpha_1 = \alpha_2 = \alpha_3 = \alpha_4 = 0.5$ (right plot) for the parameters $q = 0.9, \tau = 0.07$, and for: (a,b) $\mu/\lambda = 0, \mu/\lambda = Pi/2, \mu/\lambda = Pi$, (c,d) $\Delta/\lambda = 2, \Delta/\lambda = 4, \Delta/\lambda = 6$, (e,f) $\chi/\lambda = 0.01, \chi/\lambda = 0.1, \chi/\lambda = 0.8$

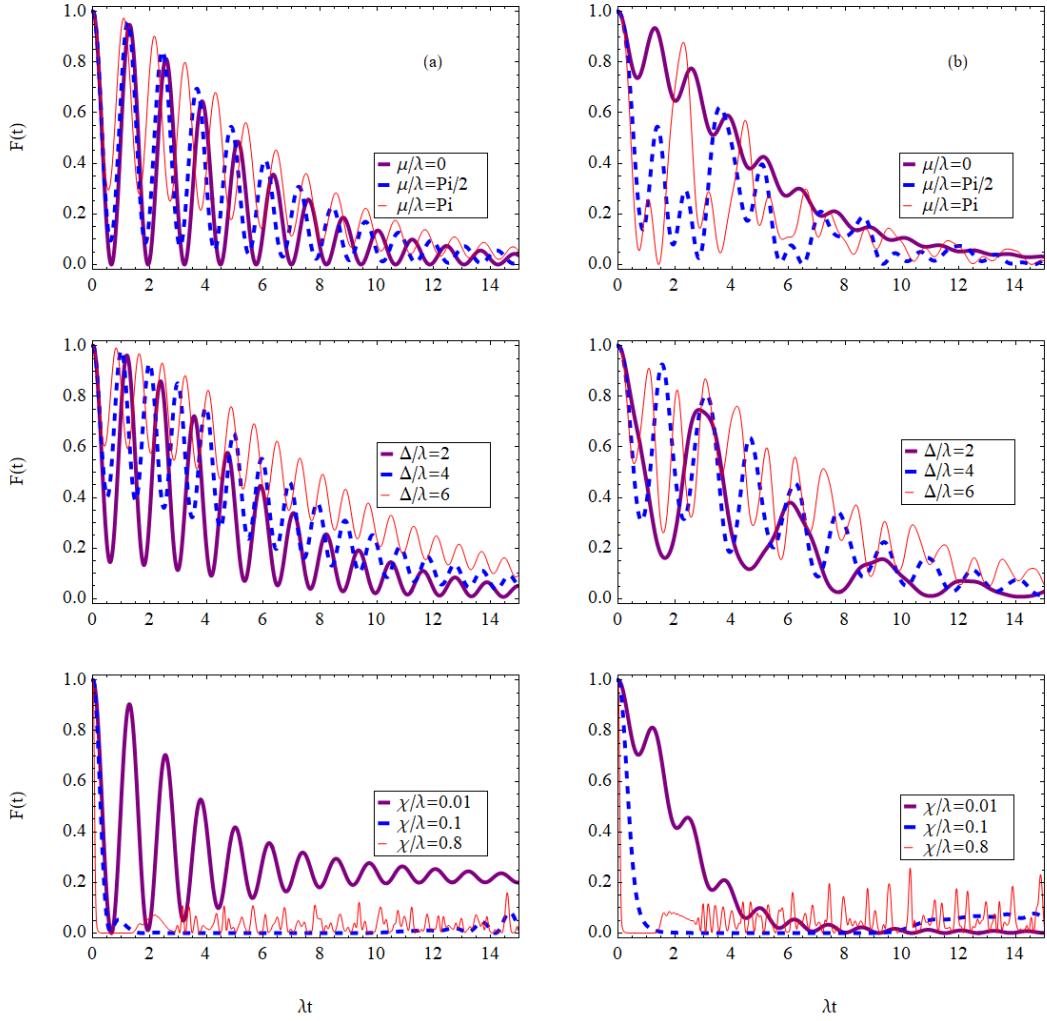

Fig. 4 Dynamics of the fidelity $F(T)$ with the same conditions as stated in Fig. 3 but for $\tau = 0.8$.

Now, we shall study the behavior of the fidelity as a function of the scaled time $T = \lambda t$ in the time-dependent/independent case. The initial field state is a q-deformed binomial state for $M = 30, q = 0.9$, and for different values of the parameter $\tau$. This will be done on the basis of the previous calculations. We examine the influence of the timedependent coupling parameter $\mu/\lambda$, detuning parameter $\Delta/\lambda$, and Kerr-medium $\chi/\lambda$ on the behavior of the fidelity for different values of the parameter $\tau$. The temporal evolution of the fidelity has given in Figs. 2-4 for $\alpha_1 = 1, \alpha_2 = \alpha_3 = \alpha_4 = 0$ in the left plot and the right plot for $\alpha_1 = \alpha_2 = \alpha_3 = \alpha_4 = 0.5$.

For a very small value of $\tau(\tau = 0.0007)$, we have plotted Fig.2. In Fig. 2(a) and Fig. 2(b), we have considered the time-dependen coupling parameter $\mu/\lambda = 0, Pi/2, Pi$ in the absence of the detuning parameter and Kerr-medium ($\Delta/\lambda = \chi/\lambda = 0$). The behavior of the fidelity started from one and exhibits the periodic collapse and revival phenomena.

When $\mu/\lambda = 0$, the time interval of the period is $\frac{2n\pi}{\sqrt{3}}, n = 1,2,3,...$ (see Fig. 1(a)). Also, as the parameter $\mu/\lambda$ incereases the number of oscillitions increases and the amplitude of oscillitions decreases. When the atom initially in a superposition of all the four levels with equal probability ($\alpha_1 = \alpha_2 = \alpha_3 = \alpha_4 = 0.5$), the behavior of the fidelity started from $F(T) = 0.125$ for $\mu/\lambda = 0, Pi/2, Pi$ (see Fig. 1( b) ). In fact, a fidelity value less than at the begining indicates that the initial state of the combined atom-field system is not perfectly aligend with expected or desired reference state being used to compute fidelity.

In Fig. 2(c) and Fig. 2(d), the behavior of the fidelity under the effect of different values of the detuning parameter is presented. The fidelity started from $F(T) = 0.064$ for $\Delta/\lambda = 2,4,6$ (see Fig. 2(d)). As we said before in Fig. 2(b), $F(T)$ started from a value less than 1 because the overlap between the two states is not perfect, implying differences in either the atom superposition state, the fields deformation, or the correlations between them. The Behavior of fidelity not affected by Kerr-medium parameter (smaller or higher values) as $\tau \to 0$.

Gradually, as $q \to 1$ and $\tau \to 1$, the q-deformed quantum effects vanish and the state behaves like a standared binomial state. We observed that the peiodicty that appeared in Fig. 2 is disappeared as $\tau$ increases and the effect of Kerr-medium on the behavior of fidelity is appeared (see Fig. 3 and Fig. 4). Also, we observed that the initial value of the fidelity when the atom initially in a superposition state is getting close one as $\tau \to 1$ (see Fig. 3(b, d) and Fig. 4(b, d) compared with Fig. 2(b, d)). The infulence of Kerr-medium on the fidelity in Fig. 4(e, f) is more prononced. For $\chi/\lambda = 0.1$ in Fig. 4(e, f), The fidelity started from one and decreases until it reaches a stable state ($F(T) = 0$). As the value of the Kerr-medium parameter $\chi/\lambda$ increased to 0.8 , a small oscilliations have apppeared with some sharp peaks.

# 4 Linear Entropy

The linear entropy of the system can be used as a tool to indicate the degree of the entanglement between the components of the system. So we devote this section to discuss the linear entropy of the system under consideration. It can be determined from the relation [17-19]

$$LE(T) = 1 - \text{Tr}\varrho_F^2(T) \qquad (4.1)$$

# References

[1] M. O. Scully, M. S. Zubairy, Quantum optics, (Cambridge University Press, 1997).

[2] C. C. Gerry, P. L. Knight, Introductory quantum optics, (Cambridge University Press, 2005).

[3] M. A. Nielsen, I. L. Chuang, Quantum computation and quantum information, (Cambridge University Press, 2000).

[4] P. A. M. Dirac, The principles of quantum mechanics, (Oxford University Press, 1930).

[5] E. T. Jaynes, F. W. Cummings, Comparison of quantum and semiclassical radiation theories with application to the beam maser, Proc. IEEE 51(1), 89-109 (1963).

[6] M. Abdel-Aty, Influence of a Kerr-like medium on the evolution of field entropy and entanglement in a three-level atom, J. Phys. B: At. Mol. Opt. Phys. 33, 2665.2676 (2000).

[7] A. Joshi, M. Xiao, On the dynamical evolution of a three-level atom with atomic motion in a lossless cavity, Optics Communications 232, pp. 273-287 (2004).

[8] C. C. Gerry, Multi-level atom-field interactions and quantum state synthesis." Phys. Rev. A 53(4), 2857.2865 ( 1996).

[9] N. H. Abdel-Wahab, The interaction between a four-level N-type atom and two-mode cavity field in the presence of a Kerr medium, J. Phys. B: At. Mol. Opt. Phys. 40(21), 4223-4233 (2007).

[10] N. H. Abdel-Wahab, A study of the interaction between a five-level atom and a single-mode cavity field: Fan-type, Modern Phys. Lett. B 25(24), 1971-1982 (2011).

[11] S. Abdel-Khalek, M. Algarni, K. Berrada, Quantum model in the context of six-level atom and deformed Lie algebra: Entanglement and statistical properties, AIP Adv. 14(3), 035343 (2024).

[12] M. Abdel-Aty, G. M. Abd Al-Kader, A.-S. F. Obada, Entropy and entanglement of an effective two-level atom interacting with two quantized field modes in squeezed displaced fock states, Chaos Solitons Fractals 12(13), 2455-2470 (2001).

[13] S. T. Korashy, A. S. Abdel-Rady, A.-N. A. Osman, Influence of Stark shift and Kerr-medium on the interaction of a two-level atom with two quantized field modes: a time-dependent system, Quantum Inf. Rev. 5(1), 9-14 (2017).

[14] A. del Rio-Lima, F. J. Poveda-Cuevas, O. Castaños, Entanglement of a three-level atom interacting with two-modes field in a cavity, J. Phys. B: At. Mol. Opt. Phys. 57, 185001 (2024).

[15] J. L. Ding, B. P. Hou, Squeezing and entanglement of a two-mode field in a four-level tripod atomic system, Opt. Commun. 284(12), 2949-2954 (2011).

[16] J. Von Neumann, Mathematical foundations of quantum mechanics, (Princeton University Press, 1932).

[17] G. Vidal, R. F. Werner, Computable measure of entanglement, Phys. Rev. A 65(3), 032314 (2002).

[18] W. J. Munro, D. F. V. James, A. G. White, P. G. Kwiat, Maximizing the entanglement of two mixed qubits, Phys. Rev. A 64, 030302 (2001).

[19] S. T. Korashy, T. M. El-Shahat, N. Habiballah, H. El-Sheikh, M. Abdel-Aty, Dynamics of nonlinear time-dependent two two-level atoms in a two-mode cavity, Int. J. Quantum Inf. 18(03), 2050003 (2020).

[20] M. S. Abdalla, E. M. Khalil, A.-S. F. Obada, Time-dependent interaction between a two-level atom and a su (1,1) Lie algebra quantum system, Int. J. Mod. Phys. B 31, 1750211 (2017).

[21] W. H. Zurek, Decoherence and the transition from quantum to classical." Phys. Today 44(10), 36-44 (1991).

[22] S. J. Anwar, M. Usman, M. Ramzan, M. K. Khan, Decoherence effects on quantum Fisher information for moving two four-level atoms in the presence of Stark effect and Kerr-like medium, Quantum Inf. 75(8), 235 (2021).

[23] H. P. Breuer, F. Petruccione, The theory of open quantum systems, (Oxford University Press, 2002).

[24] A. O. Castro, N. F. Johnson, L. Quiroga, Dynamics of quantum correlations and linear entropy in a multi-qubit-cavity system, J. Opt. B: Quantum Semiclass. Opt. 6(8), S730 (2004).

[25] E. M. Khalil, H. Abu-Zinadah, M. Y. Abd-Rabbou, Influence of an external classical field on a Ξ four-level atom inside a quantized field, Symmetry 14, 811 (2022).

[26] A. Ur Rahman, S. Haddadi, M. Javed, L. T. Kenfack, A. Ullah, Entanglement witness and linear entropy in an open system influenced by FG noise, Quantum Inf. Process. 21(11), 368 (2022).

[27] A. Ur Rahman, M. Javed, A. Ullah, Z. Ji, Probing tripartite entanglement and coherence dynamics in pure and mixed independent classical environments, Quantum Inf. Process. 20, 321 (2021).

[28] M. Akremi, S. T. Korashy, T. M. El-Shahat, R. Nekhili, Inamuddin, M. R. Gorji, I. Khan, New features of non-linear timedependent two-level atoms, J. Taiwan Inst. Chem. Eng. 105, 171.181 (2019).

[29] S. Korashy, M. Abdel-Aty, Quantum control of a nonlinear time-dependent interaction of a damped three-level atom, axioms 12, 552 (2023).

[30] N. Zidan, S. Abdel-Khalek, M. Abdel-Aty, Geometric phase and disentanglement of a moving four-level atom in the presence of nonlinear medium, Int. J. Quantum Inf. 10, 1250007 (2012).

[31] S. T. Korashy, Mohammed A. Saleem, T. M. El-Shahat, Geometric phase of a moving four-level lambda-type atom in a dissipative cavity, Appl. Maths. Info. Sci. 15(2), 153-163 (2021).

[32] S. T. Korashy, H. Al Bayatti, T. M. El-Shahat, Statistical properties of the nonlinear time-dependent interaction between a three-level atom and optical Fields, J. Stat. Appl. Pro. 10(3), 779-793 (2021)

[33] P. M. Poggi, G. De Chiara, S. Campbell, A. Kiely, Universally robust quantum control, Phys. Rev. Lett. 132, 19380 (2024).

[34] I. Khalid, C. A. Weidner, E. A. Jonckheere, S. G. Shermer, F. C. Langbein, Statistically characterising robustness and .delity of quantum controls and quantum control algorithms. Phys. Rev. A 107, 032606 (2023).

[35] D. Stoler, B. E. A. Saleh, M. C. Teich, Binomial state of the quantized radiation .eld, J. Mod. Opt. 32(3), 345.355 (1985).

[36] G. Dattoli, J. Gallardo, A. Torre, Binomial state of the quantized radiation .eld: Comment, J. Opt. Soc. Am. B 4(2), 185.191 (1987).

[37] J. R. Klauder, B. Skagerstam, Coherent states, (World Scienti.c, Singapore, 1985).

[38] J. P. Gazeau, Coherent states in quantum physics, (Wiley-VCH, Berlin, 2009).

[39] S. C. Jing, H. Y. Fan, q-deformed binomial state, Phys. Rev. A. 49(4), 2277-2279 (1994).

[40] S. C. Jing, The q-deformed binomial distribution and its asymptotic behaviour, J. Phys. A: Math. Gen. 27, 493-499 (1994).

[41] C. Quesne, New q-deformed coherent states with an explicitly known resolution of unity, J. Phys. A: Math. Gen. 35 , 9213-9226 (2002).

[42] A. J. Macfarlane, On q-analogues of the quantum harmonic oscillator and the quantum group SU(2)q, J. Phys. A: Math. Gen. 22, 4581 (1989).

[43] L. C. Biedenharn, The quantum group SUq (2) and a q-analogue of the boson operators, J. Phys. A: Math. Gen. 22(18), L873 (1989).

[44] V. G. Drinfeld, Quantum groups, Proc. ICM-86 (Berkeley) 1, 798-820 (1987).

[45] F. H. Jackson, On Q-de.nite integrals, Q. J. Pure Appl. Math. 41, 193-203 (1910).

[46] S. C. Jing, H. Y. Fan, A new completeness relation in the q-deformed two-mode Fock

space, J. Phys. A: Math. Gen. 26, L69 (1993).

[47] H. Y. Fan, S. C. Jing, Integration within an ordered product for q-deformed bosons, Phys. Lett. A 179(6), 379-384 (1993).

- E-mail address: korashy.quantum2025@gmail.com